\documentclass[%
 reprint,
 amsmath,amssymb,
 aps,
]{revtex4-2}

\usepackage{graphicx}
\usepackage{dcolumn}
\usepackage{bm}

\begin{document}

\preprint{APS/123-QED}

\title{Influences on Faculty Uptake from a Faculty Learning Community}

\author{Lydia G. Bender}
\affiliation{%
 Department of Physics, Kansas State University, Manhattan, Kansas 66506\\
}%

\author{James T. Laverty}
\affiliation{%
 Department of Physics, Kansas State University, Manhattan, Kansas 66506\\
 \email{\textit{Correspondence to:} laverty@ksu.edu}
}%


\date{\today}

\begin{abstract}
Professional development is a tool that faculty members use to develop knowledge and skills that help them become better teachers. We investigate what influences affect the ways in which faculty take up ideas from professional development programs. By employing the framework of Pedagogical Reasoning and Action, we investigate how faculty take up ideas from a particular Faculty Learning Community (the STEM Teaching \& Learning Fellowship) and the factors that influence their instructional and material design choices. We examined influences affecting faculty in three different cases. From these cases, we constructed themes, and examined those themes across all cases using a cross-case analysis. In this multiple case study we find that assessment and instructional alignment, participation in instructional practices, and the culture of the department all influence the extent in which faculty bring new teaching ideas and practices into the classroom. These findings can be leveraged by developers to design and improve programs, as well as inform researchers on future avenues of research.



\end{abstract}

\maketitle


\section{Introduction}
Efforts to improve STEM education focus on increasing retention rates and preparing students for careers in STEM~\cite{mohr2019, national2012framework, olson2012}. Teachers at the post-secondary level play a large role in increasing STEM literacy among their students. As new practices and pedagogies are created to improve student outcomes, teachers learn and apply these techniques to their classrooms. One way that faculty can develop new skills, techniques, and knowledge is through professional development~\cite{guskey2000}.
Research on professional development includes, but is not limited to, the design of programs by creating observational tools~\cite{olmstead2016}, long-term sustainability of programs~\cite{tinnell2019, sirum2010}, correlation between adoption of professional development principles and student outcomes~\cite{elliott2016}, and faculty use of instructional practices~\cite{ebert-may2011, manduca2017}. Review of the literature also reveals a desire to conduct research about faculty during their participation in programs~\cite{southerland2016, daly2011, henderson2011facilitating}. 

This paper examines faculty members participating in a STEM Teaching \& Learning Fellowship, henceforth referred to as the Fellowship. The Fellowship is part of a larger project that involves bringing Three-Dimensional Learning (3DL) to undergraduate physics classes. Three-Dimensional Learning is the foundation of the Next Generation Science Standards \cite{national2012framework} and is largely focused on integrating scientific practices, crosscutting concepts, and core ideas into the classroom. Integration of these dimensions into the classroom helps create scientific literate citizens that possess skills needed to start their careers \cite{national2012framework}.

Other research that has been conducted by this group has focused on developing Three-Dimensional instruction and assessment protocols~\cite{bain2020, laverty2016}, the 
alignment between 3DL and common concept inventories~\cite{laverty2018}, 3DL assessment interventions~\cite{underwood2018}, 3DL course transformations~\cite{matz2018}, and explanations of specific principles of 3DL~\cite{cooper2020crosscutting}. The Fellowship provides us with the space to investigate different avenues of research within the same context including research on the program itself.

The Fellowship focuses on bringing Three-Dimensional Learning into undergraduate STEM classrooms in order to provide students with meaningful and transferable knowledge. Our research is focused on physics faculty that are asked to design material for a unit of their course using ideas from the Fellowship. Throughout the process of material development the 3DL framework is used as a way to establish a common language for discussing and aligning learning goals, assessment, and instruction. The material that they create and the process in which that occurs allows us to investigate how faculty take up ideas from the Fellowship and apply it to their classrooms.

Moreover, our research views faculty as talented educators with good ideas and the appropriate skills to implement them. Because we want faculty to carry out their ideas regarding their material design, it is essential to investigate the influences that affect their design and use. In addition, by identifying influences that affect the ways in which faculty take up ideas from a program we are able to pinpoint features of program design that support faculty in implementing new ideas and practices.

Through the course of this paper, we review the literature pertaining to STEM professional development in higher education, faculty learning communities, and faculty uptake in Section~\ref{litrev}. In Section~\ref{context}, we explore the structure and design of the STEM Teaching \& Learning Fellowship before arriving at our research questions in Section~\ref{resquest}. In Section~\ref{PRA}, we describe how we employed Pedagogical Reasoning and Action~\cite{shulman1987} to investigate influences on the design choices of faculty. Using case studies, we examine three faculty members, Ron, Charlie, and Molly (pseudonyms), and present three themes that were developed using a cross-case analysis. Details of the cross-case analysis is outlined in Section~\ref{methods} and the results of the analysis can be found in Section \ref{cases}. In Sections~\ref{discussion} \&~\ref{conclusion} we go on to discuss our findings in the context of other research as well as our recommendations for future development of faculty learning communities.

\section{Literature Review\label{litrev}}

\subsection{Approaches to Professional Development\label{approaches}}

Professional development programs are designed to evoke change. Henderson, Beach, and Finkelstein categorized professional development efforts by conducting a meta-analysis of 191 journal articles that focused on promoting change~\cite{henderson2011facilitating} in higher education STEM. Review of these articles revealed four change categories that can be distinguished along two axes. One axis describes what the program intends to change (individuals vs. environments and structures) while the other axis describes who has control over the purpose of the program (prescribed vs. emergent). Professional development that falls under “Individual” focuses on instructors' beliefs and behaviors~\cite{henderson2011facilitating}, and professional development that belongs to “Environment” focuses on changing environments and structures that influence instructional choices~\cite{henderson2011facilitating}. Programs considered “Prescribed” are led by one or more individuals separate from the participants that come into the program to teach a predetermined set of ideas. “Emergent” programs do not begin with predetermined set of ideas. Instead, the content and delivery are driven by the participants' needs. 

Henderson, Beach, and Finkelstein find that emergent strategies that align with the change beliefs of instructors over an extended amount of time tend to be most successful~\cite{henderson2011facilitating}. These approaches align with an asset-based approach to professional development which views faculty as experts of their own local contexts, knowledge, values, and tools~\cite{missingham2017, strubbe2020}. It also celebrates and builds upon the ideas that come from a diverse group of faculty participants ~\cite{swan2017}. These approaches also seek to leverage the ideas and skills of faculty members and can ultimately develop faculty to become change agents~\cite{missingham2017,kezar2013, strubbe2020}. One way that professional development can take an asset-based approach is through the use of Faculty Learning Communities.

\subsection{Faculty Learning Communities\label{flc}}
Faculty Learning Communities (FLCs) are one type of professional development that incorporates different levels of faculty participation and ownership~\cite{daly2011,richlin2004} through the use of a community of faculty who support each other~\cite{natkin2016}. FLCs occur over extended amounts of time, where faculty and facilitators come together to investigate and discuss different teaching practices and concepts~\cite{natkin2016, cox2004,layne2002}. They also provide faculty with the space and time to reflect on their teaching~\cite{petrone2004,layne2002}. 

One common feature of FLCs is that participants often have varying levels of teaching experience and come from different disciplinary backgrounds~\cite{cox2004, wicks2015}. These features have the ability to instill confidence in younger colleagues~\cite{layne2002}, and the interdisciplinary feature of FLCs can provide faculty with the opportunity to interact with new colleagues and talk about similar problems and share their different approaches to solving those problems~\cite{layne2002}. The FLC also gives them the platform to develop a community of colleagues with whom they can discuss professional and personal topics outside of the FLC~\cite{layne2002}.

More formal FLCs are often led by facilitators that are focused on participants' ideas and needs. The role of the facilitator is to create a productive space for the community, keep a focus on the big picture, and practice organizational skills~\cite{petrone2004}. The facilitator can also take on the responsibility of training faculty to use a certain tool or resource~\cite{Nugent2008}.

FLCs take many different forms in order to account for the different types of faculty participating within them. Two common types of FLCs are Faculty Online Learning Communities (FOLC)~\cite{dancy2019, rundquist2015} and University-Affiliated Faculty Learning Communities (UFLC)~\cite{glowacki2006}. FOLCs occur online and can extend over multiple institutions, this allows for a more narrow focus on professional development because faculty who are interested in a specific topic can join the community virtually~\cite{dancy2019}. UFLCs are composed of faculty that all come from the same university, and tend to participate in a more structured way than an independent learning community.

\subsection{Faculty Uptake of Pedagogical Practices\label{uptake}}
Research has focused on faculty use of practices and skills after participation in professional development programs. Some research has looked into how faculty report on their use of materials~\cite{ebert-may2011, d2008, kane2002}, others look at influences of discontinuation of use after participation in professional development~\cite{henderson2012, stains2015, dancy2012, henderson2007}, and others investigate faculty’s selection and use of new practices ~\cite{zohrabi2020, turpen2016}.

Henderson, Dancy, and Niewiadomska-Bugaj conducted a study on the factors that influence faculty to continue or discontinue use of research-based instructional strategies (RBIS) ~\cite{henderson2012}. Their research found that professional development is effective at making faculty aware of current RBIS~\cite{henderson2012}. This study also identified factors such as attending teaching related professional development, reading teaching journals, interest in RBIS, and being female have a positive effect on the continued use of RBIS ~\cite{henderson2012}. Other studies have found that the lack of local support ~\cite{dancy2012, stains2015}, student responses~\cite{dancy2012}, and time~\cite{dancy2012,sharma2010} have a negative effect on the adoption and continued use of new teaching practices. Research of this nature has suggested professional development programs should provide support and feedback during implementation ~\cite{henderson2012, blume2010}, and spend time addressing situational barriers~\cite{henderson2007}.

Other studies have also looked at the different ways in which faculty decide to use and choose to continue with new pedagogical practices ~\cite{zohrabi2020, turpen2016}. Zohrabi finds that collecting student feedback and intuition benefit faculty in their continued use of new ideas while departmental and classroom practices, student engagement, and the use of classroom materials are all factors that influence how new ideas are applied to their teaching ~\cite{zohrabi2020}. Overall, this study finds that faculty view the process of implementing pedagogical change as a positive experience that becomes easier over time ~\cite{zohrabi2020}. Another study conducted by Turpen, Dancy, and Henderson found that factors such as using personal experience to gauge effectiveness and encouragement from their community and department influence adoption and continued use of a new pedagogical practice ~\cite{turpen2016}.

\section{STEM Teaching \& Learning Fellowship\label{context}}

\begin{table}[htb]
\caption{Description of each Cohort in the STEM Teaching \& Learning Fellowship. Each Cohort consists of Fellows from all four disciplines (Physics, Mathematics, Chemistry, and Biology)\label{tab:Cohorts}}
\begin{tabular}{c|c|c|c}
\hline \hline
Cohort & Year        & Participants & Institutions         \\ \hline
1      & 2014 - 2016 & 9 Fellows    & Single site                 \\
2      & 2016 - 2018 & 14 Fellows   & Single site                 \\
3      & 2018 - 2020 & 20 Fellows   & Multiple sites \\
4      & 2019 - 2021 & 12 Fellows   & Multiple sites  \\  \hline \hline  
\end{tabular}
\end{table}

Our research focuses on physics faculty participating in the STEM Teaching \& Learning Fellowship. The Fellowship started at Michigan State University (MSU) in 2013, but has since expanded to include three more universities: Florida International University (FIU), Grand Valley State University (GVSU), and Kansas State University (KSU). Since 2013 the Fellowship has had four Cohorts of faculty, each with approximately 10-20 Fellows. The first two Cohorts included faculty only at MSU, and the following Cohorts consisted of faculty from all four institutions. Although our research only focuses on physics participants, each Cohort had faculty participate from physics, chemistry, biology, and math departments. A description of each Cohort and the participating faculty can be seen in Table~\ref{tab:Cohorts}.

The principles of the Fellowship fit closely with faculty learning communities. Although the Fellowship is led by a group of Three-Dimensional Learning (3DL) experts from all four institutions, the topics are selected by faculty needs, and breakout discussions are led by the Fellows. A large portion of the Fellowship is focused on giving Fellows the time and space to reflect on their teaching. Participating Fellows met virtually once a month for ninety minutes over the course of two academic years, and this amounts to approximately 16 meetings and 24 hours of professional development.

The Fellowship is centered around the goal of integrating 3DL instruction into undergraduate STEM classes. 3DL focuses on three main components: core ideas, scientific practices, and crosscutting concepts~\cite{national2012framework}. Core ideas are fundamental ideas that are unique to each STEM discipline. In the case of physics these would include but are not limited to the ideas of energy conservation, forces and interactions, or waves \cite{national2012framework}. Unlike core ideas, scientific practices and crosscutting concepts are not unique to specific disciplines. Scientific practices are the different ways that scientists engage in scientific problem solving, and crosscutting concepts are lenses used by scientists across disciplines to inform how they engage in scientific practice \cite{cooper2020crosscutting}. The ideas of 3DL are explored through readings, presentations, and design techniques. Within the Fellowship the ideas of 3DL are used as a way for the Fellows to think about and talk about their learning goals and not as a teaching method that they must integrate into their course.

The Fellowship promotes Learning Goals Driven Design (LGDD) as the approach faculty should use to design their courses.  The principles of LGDD are modeled after Wiggins \& McTighe’s model of Backward Design~\cite{wiggins1998}. LGDD is a process of designing materials that focuses on three stages: articulating learning objectives, creating the material, and collecting feedback~\cite{krajcik2008}. More specifically, LGDD is used in the Fellowship as a tool to have Fellows integrate 3DL into their lessons by articulating their desired learning outcomes blended with 3DL, the evidence needed to reach those outcomes, and how the outcomes would be assessed.
 
Discussions and activities presented in the Fellowship are centered around the idea that assessment drives changes to instruction~\cite{laverty2014}. In order to change assessment and align that to instruction, the LGDD model is presented by facilitators and discussed by Fellows in meetings and in the forum. As part of learning about LGDD, a significant amount of time and attention is given to creating learning objectives and assessment tasks that align with the 3DL framework. Several meetings are spent discussing how to articulate what it is that instructors want their students to be able to do, and how those objectives would be assessed. After creating 3DL learning outcomes, time is spent creating assessment items that can produce evidence of achieved learning goals, and instructional materials that align with those assessment items. More specifically time is spent reflecting on their own assessment practices and utilizing the Three-Dimensional Learning Assessment Protocol~\cite{laverty2016} to align exam problems with 3DL.

During the Fellowship, Fellows are asked to participate in discussions, activities, and reflections. One of the largest tasks given to Fellows asks them to design materials for their course and use them in their classroom, using the principles of LGDD. The scope of the 3DL unit is left for the Fellows to decide, but the unit contains learning objectives, classroom activities/materials, and an assessment task. The application and design of the 3DL unit serves as the focus of our research.

\section{Research Questions\label{resquest}}
Our research is concerned with the ways that faculty take up and implement their ideas from the Fellowship into their classrooms. The first opportunity that the Fellows have to implement ideas into their classrooms is with their 3DL unit. Therefore, we are focused on the ways in which Fellows develop and implement their 3DL unit. 

The factors that we are interested in consist of tools, resources, interactions, or ideas that Fellow’s mention as part of their design process. Henderson, Dancy, and Niewiadomska-Bugaj found that most faculty stop the use of RBIS after initial implementation~\cite{henderson2012}. Therefore, it is important for us to look at Fellow’s outlook and plans for continued use because it has already been seen in research to affect uptake. 

In particular, we are interested in the following research questions:

\begin{enumerate}
    \item Which aspects of the material design and instruction process promote or hinder faculty’s intentions to continue using 3DL?
    \item How does the FLC support faculty in their adoption and plan for the continued use of 3DL?
\end{enumerate}

Exploration of these questions allows us to look at common mechanisms that influence the ways in which our faculty design and implement their 3DL material into the classroom as well as their plans for future use. Additionally, it allows us to explore the design process of faculty involved in different departmental, institutional, and social contexts. Identification of these mechanisms allows us to look at design features of FLCs, and the ways in which they can be used to support faculty.

\section{Theoretical Framework\label{PRA}}
For this study we used Pedagogical Reasoning and Action (PR\&A) to investigate how faculty approach the design of their material. PR\&A is a framework proposed by Shulman to explain a teacher’s development of classroom material and instruction from their content knowledge~\cite{pella2015}.    

PR\&A is particularly useful for faculty discussing their thoughts behind design decisions.  Shulman (1987) states, “The following conception of pedagogical reasoning and action is taken from the point of view of the teacher, who is presented with the challenge of taking what he or she understands and making it ready for effective instruction"~\cite{shulman1987}. Stroupe takes on the following definition of PR\&A, “the purposeful coordination of ideas, information, and values about subject matter, curriculum, learners, and instructional context to plan for, enact, and reflect on instructional practice"~\cite{stroupe2016beginning}. For our research we also take on this definition as it will help us examine what influences affect the design choices of our Fellow's 3DL unit. Research that has used this framework investigates the choices of beginning teachers~\cite{stroupe2016beginning}, the significance of content knowledge~\cite{loewenberg2008}, and uses/or choices of technology~\cite{niess2017,starkey2010, feng2005, smart2015, smart2013}.

PR\&A places significance on the process of thinking about instruction and not solely on the observable acts of teaching~\cite{shulman1987}. Thus, we can explore material design from the point of view of a faculty member, focusing on their talents, ideas, and practices. Furthermore, PR\&A allows us to examine their reasoning as an educator and as a subject matter specialist~\cite{endacott2015}.  

PR\&A occurs in five stages and ends with faculty creating new comprehension from their experience creating and applying their material in the classroom~\cite{shulman1987}. These stages include comprehension, transformation, instruction, evaluation, and reflection, but these stages do not necessarily progress in this order~\cite{shulman1987}. A figure of PR\&A with its stages and descriptions can be found below in Figure~\ref{fig:PRA}.

\begin{figure}[htb]
\centering
\includegraphics[scale=0.33]{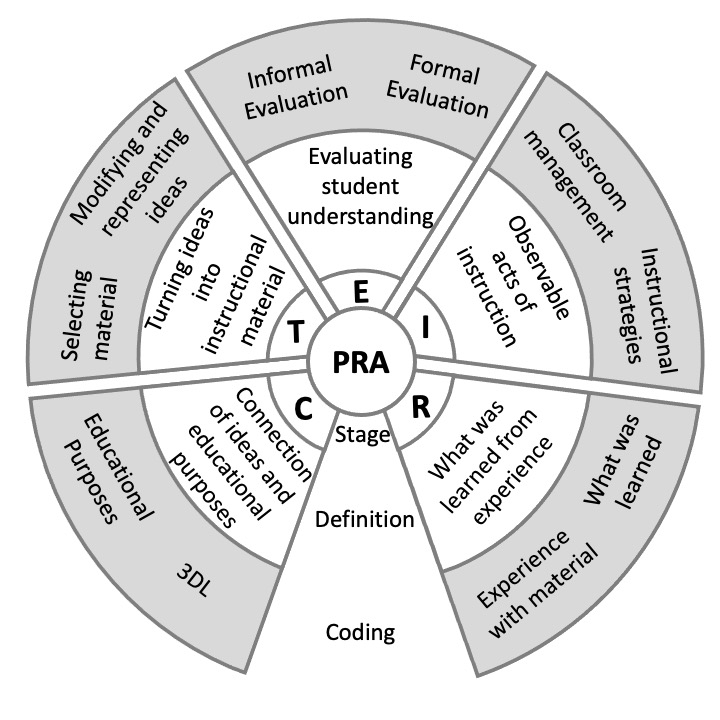}
\caption{The 5 stages of Pedagogical Reasoning and Action as well as each stage's definition and codes identified in this study}
\label{fig:PRA}
\end{figure}

\emph{Comprehension} involves faculty thinking about the set of ideas they want to teach, and how those ideas connect to the educational purposes of the class~\cite{shulman1987}. Comprehension is attained “when teachers understand what they are going to teach~\cite{smart2013}." As experts in their fields our faculty are knowledgeable in physics content; therefore, in our research we focused on their comprehension of 3DL, and the ways that it intersects with the educational purposes of their class.  The information that is learned from their experience implementing the material forms their \emph{New Comprehension}, and then becomes part of their comprehension base as they move forward creating new materials.  

In this study, the \emph{Transformation} stage broadly focuses on how faculty turn their 3D design ideas into instructional material. This occurs in five substages: preparation, representation, instructional selections, adaptation, and tailoring~\cite{shulman1987}.  These five substages broadly focus on selecting what is to be taught, representation of ideas, making sure materials are in an instructional format, and modifying ideas so that they are suitable for the students in their classroom. Ultimately this “result[s] in a plan, or set of strategies, to present a lesson, unit, or course”~\cite{shulman1987}.

\emph{Evaluation} is centered around ways faculty choose to evaluate student understanding inside of the classroom as well as formally testing their understanding. Inside of the classroom, faculty may choose to use clicker questions or reflections to test student understanding of material during class time whereas formal evaluation may look like end of unit exams.

The next stage of PR\&A is \emph{Instruction}; this stage focuses on the observable acts of instruction~\cite{shulman1987}.  Examples of these observable acts include classroom management, presentation of material, and interactions with students~\cite{smart2013}. In our research, video data of the 3DL unit was not available; therefore, we were unable to observe instruction of Fellow's 3DL unit. In lieu of video data, faculty were asked to recall and describe their instruction of the unit as part of an interview. 

The \emph{Reflection} stage is focused on but is not limited to recalling impressions and feelings of their teaching and overall experience of their material~\cite{smart2013}. In our study we have Fellows reflect on how they felt their 3DL unit went, and what they learned from implementation. Fellow’s reflection on their unit allows us to determine their intentions for future use of their materials.

\section{Methodology\label{methods}}
In this paper, we use a case study approach, which allows us to deeply explore a phenomenon within its context~\cite{creswell2007}. This is typically done through the use of multiple sources of data in order to produce case-based themes~\cite{creswell2007}. For our study, a case is confined to the experience of a Fellow participating in the Fellowship. By exploring each case, we are able to gain a thorough insight into their participation in the Fellowship, design of their 3DL unit, the Fellowship itself, and the ways in which these aspects influence their PR\&A and hence their design choices.

In order to explore the different ways in which faculty take up ideas from the Fellowship we selected three different cases. These cases are then cross analyzed to complete a multiple case study. A multiple case study gives us the opportunity to explore themes, reoccurring ideas and patterns, in more than one Fellow’s experience designing material. Moreover, comparing multiple cases and their case-based themes together allows us to explore both the similarities and differences in 3DL material design and the underlying mechanisms that influence those choices.

\subsection{Case Selection}

We chose to explore the experiences of three physics Fellows: Ron, Charlie, and Molly. We chose to focus on physics Fellows because our physics expertise would lend itself to understanding the design of physics classroom materials as well as the physical and cultural contexts that our Fellows participate in. We also chose these cases based on our informal understanding of their attitudes toward 3DL and the Fellowship. Our cases encompass a range of attitudes. Our first Fellow Ron was excited about 3DL, our second Fellow Charlie felt skeptical about it, and our final Fellow Molly was familiar with 3DL as she already uses principles of 3DL in her teaching. Our cases are also composed of faculty with different levels of experience teaching and teach in different classroom formats.  Finally, these Fellows also use a range of different tools and resources that affect the way that they participate in different categories of PR\&A. A more comprehensive background of each of the Fellows can be found in Table~\ref{tab:cases}.

\begin{table*}[htb]
\caption{A summary of the three selected cases \label{tab:cases}}
\begin{tabular}{l|c|l|l|l}
\hline \hline
        & Cohort & Academic Level & Classroom Format                  & Attitude Towards 3DL           \\ \hline
Ron     & Later cohort      & Pre-Tenure     & Lecture                           & Excited about 3DL              \\ 
Charlie & Later cohort      & Post-Tenure    & Lecture, Lab, Recitation      & Skeptical about the use of 3DL \\
Molly   & Earlier cohort      & Pre-Tenure     & Studio  & Already uses principles of 3DL \\
\hline \hline
\end{tabular}
\end{table*}

\subsection{Data Collection}
In order to gain a deep understanding of our Fellows, we collected data from interviews, Fellowship assignments, Fellowship meetings, and forum discussions. 

The interview was semi-structured, lasted for approximately 45 minutes, and occurred either online or in person. Ron and Charlie were interviewed after their first year in the Fellowship and Molly was interviewed one year after she completed the Fellowship. The interview was focused on the design of 3D material and was constructed to specifically address all five stages of PR\&A. The interview provided Fellows with an opportunity to vocalize their understanding and thoughts about 3DL, their experience in the Fellowship, and the choices they make when designing material. In order to anchor the conversation around their 3DL material, we used stimulated recall techniques~\cite{gass2000} by introducing the material they designed during the Fellowship into the interview. 

The Fellowship assignments we collected were posted to a faculty forum and included an end of the year survey, reading reflections, their 3DL unit, and other prompts that Fellows were asked to respond to as part of the Fellowship. Earlier cohorts, like the one that Molly participated in, did not participate in forum discussions or end of year surveys, but they did submit their 3DL material. For later cohorts, the end of the year survey focused on each Fellow’s personal goals and accomplishments. It asked them to talk about how the Fellowship supported their use of 3DL, and the supports that they needed. As well as collecting assignments, the discussions that occurred in the comment section of the forum were also collected. 

The meeting recordings we collected were from disciplinary Fellowship meetings from the later cohorts. Fellows that attended these meetings were from different institutions but belonged to the same STEM discipline. These meetings focused on each Fellows 3DL unit and served as a way for Fellows to share ideas and receive feedback in a smaller, more intentional way.

\subsection{Data Analysis}
\begin{figure*}[htb]
\centering
\includegraphics[scale=0.19]{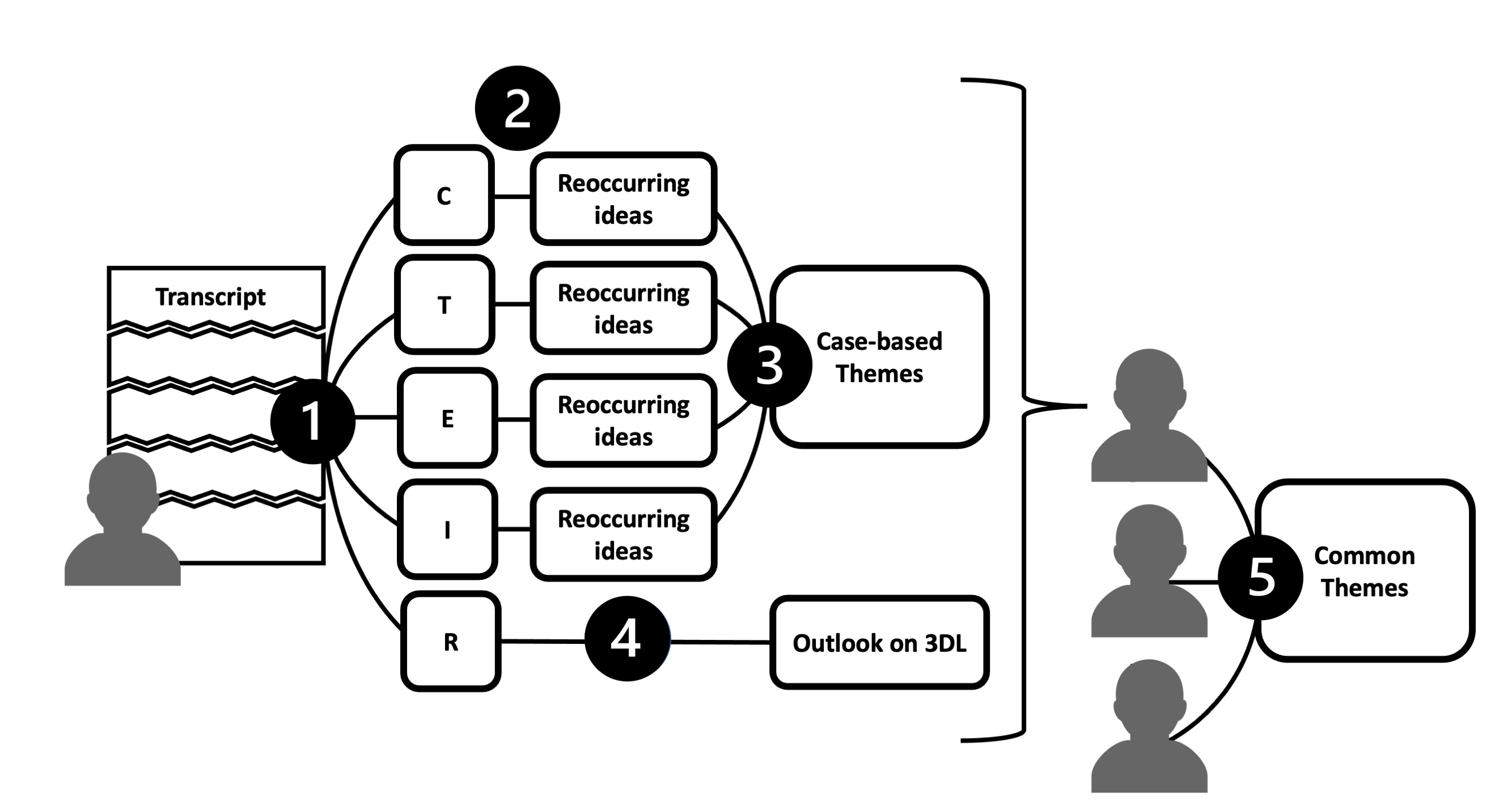}\caption{
The data analysis started by 1) identifying pieces of data that relate to the material design process and sorting them into the five stages of PR\&A, then 2) identifying reoccurring ideas within each of those stages.  The reoccurring ideas for each subject were then 3) compared across the first four stages to synthesize case-specific themes. Those themes were then 4) compared to the faculty members outlook on continuing to use 3DL and then 5) all of these connections were synthesized to determine our  results.
}
\label{fig:DA}
\end{figure*}

Data analysis occurred in five separate stages: sorting and coding, theme development, synthesis, linking to outlook, and cross-case analysis. The first four stages occurred within a single case, and the last step occurred during the cross-case analysis between all three cases. The process of data analysis is summarized in Figure~\ref{fig:DA} and is described in detail below. 

Data analysis started by sorting and coding our data into each stage of PR\&A. Coding is a process that allows us to make sense of the data by “indexing or mapping data, to provide an overview of disparate data that allows the researcher to make sense of them in relationship to their research questions~\cite{elliott2018}.” In order to sort and code the data we looked at each Fellow separately. We pulled every response and comment that Fellows made within the data set to be further analyzed. Data was separated by individual response. For verbal data, individual responses were sorted by their responses to individual questions and/or individual comments made, and the written data were split by into their paragraphs. After the data was sorted by individual response, they were read and coded into a stage of PR\&A. 

We sorted our data into a single stage of PR\&A by looking at the main idea that is being conveyed~\cite{creswell2020}. For example, if the main idea of the statement was to articulate the educational purpose of their class then it would be sorted into the comprehension stage. Data that did not belong to a stage of PR\&A were not thrown out but were instead used to provide more details about the case (e.g. motivation to join the Fellowship). Inter-rater reliability was conducted with our data coding. Initially researchers came to an agreement of 82\%, after discussions between researcher consensus between the researchers increased to 96\%. 

After the data was sorted and coded into each stage of PR\&A we looked for themes within each stage of PR\&A. Theme development occurs by looking through the coded data and extracting common ideas that brings the data together~\cite{nowell2017}. During the theme development of our research, we looked at each category and paid close attention to features of their context or the Fellowship that Fellows note as helping or hindering them. Factors or ideas that were mentioned multiple times or in detail were recorded and were used to compose our themes for that stage of PR\&A. 

Once each stage of PR\&A was considered individually, we examined the comprehension, transformation, evaluation, and instruction stages to find reoccurring ideas that affected multiple parts of a Fellow’s PR\&A. In doing this we were able to focus on influences that occurred in several stages of PR\&A, and therefore were influencing large parts of the Fellow’s process of material design. The reflection stage was examined in order to find each Fellow’s outlook on 3DL. Plans for continued use and feelings about 3DL are considered parts of outlook on 3DL. 

The next part of our analysis involved linking the themes to the Fellow’s outlook of 3DL and their material. Their outlook of 3DL includes their plans for continued use or outcome of the 3DL unit.  This step was conducted in order to create a deeper understanding on how the themes affect our Fellow’s 3DL material design and continued use of 3DL. 

Our last step of data analysis was a cross-case analysis. During this analysis we compared and contrasted themes between Fellows. In particular we looked for similar experiences and instances that the Fellows had and how this affected their material design, instruction, and continued use of 3DL. This allowed us to identify and explore the underlying mechanisms influencing material design.

\subsection{Limitations}
The limitations present in our study come from our focus on physics, our case selections, and our use of the PR\&A framework. 

The Fellowship is interdisciplinary, but we chose to focus on physics Fellows because our physics expertise would lend itself to understanding the design of physics classroom materials as well as the physical and cultural contexts that our Fellows participate in.  Since this study was conducted with physics faculty, the claims that are made are confined to physics departments and professional development programs. Although it would be reasonable to assume that course structures and alignment of assessment and instructional materials are all factors that would influence material design for other STEM Fellows outside of the physics discipline.

The Fellows that we selected for our cases are all faculty members at large and predominantly white institutions. They also all freely chose to participate in a long-term teaching Fellowship. The demographic make-up of the institutions that our Fellows participate in affects the culture of their department as well as the considerations that Fellows take into account when designing classroom material. These considerations could look vastly different for Fellows that do not work at similar universities; therefore, our claims about the culture of departments are limited to Faculty participating in large and predominately white universities. Our Fellows also chose to participate in a two year teaching Fellowship because of the practices that our Fellows use may not extend to the experiences that every faculty member faces when designing instructional and classroom materials.

The use of the framework of Pedagogical Reasoning and Action also limits the scope of our research investigation. By focusing on Fellows’ PR\&A we were able to investigate the influences on the design of a Fellows’ classroom material, but not influences that affect other experiences that our Fellows face inside of the Fellowship. These experiences may include their motivation to join the Fellowship, new ideas learned that go beyond instruction and material design, and interactions that influence other areas of their professional life.

\section{Results\label{cases}}

Ron, Charlie, and Molly teach introductory physics courses, but have different student populations. They vary in their teaching experience, motivations to join the Fellowship, and material implementation. Ron is a pre-tenure professor teaching a lecture-only introductory physics course, Charlie is tenured teaching an introductory physics course for non-physics science majors, and Molly is a pre-tenure faculty member who teaches a studio-based (integrated lab and lecture) physics course taken primarily by biology majors.  

Each one of the Fellows were motivated to join the Fellowship for different reasons. Ron was motivated to join to “meet like-minded people” and to learn how to prepare lectures using “evidence-based approaches.” Molly and Charlie were motivated to join in order to improve their classes. Charlie tries to find new ways he can improve his teaching every couple of semesters. Molly on the other hand teaches a class that is focused on giving students the space to explore physics phenomena through the use of experiments and because of this, Molly wanted to learn how to assess her students on these lab skills.


In order to look for themes that influence Fellow's design and implementation of 3DL material, we cross-analyzed each case by comparing and contrasting reoccurring ideas from each one of the four stages of a Fellows’ PR\&A and the ways in which they affected the Fellows outlook on 3DL. Our data analysis revealed three themes that were influencing all three of our Fellows. These themes include alignment of assessment and instructional practices, material placement and participation in instruction, and engagement in social interactions. 


Below we present our themes and how they connect to each Fellow’s outlook on 3DL with supporting evidence from each Fellow’s case. Because we used multiple sources of data, we distinguish them by using~[I] to signify an interview,~[M] for a Fellowship or disciplinary meeting, and~[F] for forum discussions and homework assignments.

\subsection{Alignment of Assessment and Instructional Practices}
The first theme developed from the data was the motivation/ability to change and align assessment and instructional practices. This theme arose in the comprehension and evaluation stages of our Fellows’ PR\&A and is ultimately connected to Fellow’s reservations about the use of 3DL. 

\subsubsection{Comprehension}

All three of our Fellows have educational goals for their classrooms. These goals are related with 3DL and focus on assisting students develop scientific and critical thinking skills that they can apply to different areas in their education and life.  

Ron’s learning goal for his class “is to help them learn problem solving skills~[I].” More specifically Ron wanted his “students to do [every scientific practice] in the classroom, […] These are the practices that I'd like to see my students do~[I].” These learning goals also extend to students in his advanced lab course. When talking about his advanced lab course Ron states, “I mean, there's just so much of 3D that you should be doing~[I].” Overall, Ron believes the principles of 3DL, specifically how it teaches scientific practices, aligns well with what he wants students to learn.

Charlie has multiple educational goals that align with 3DL principles. First, Charlie wants students to “try and you know, make that connection between the physics […] and the scientific field that most of my students are interested in~[I].” He would also like his students to “have a feel for, you know, a little bit of a physical intuition about the world.~[I]” He also wants his students to be able to synthesize information from different areas and bring them together to solve a problem. Charlie states, “it's a physics class and so we do you know, there's a lot of problems worked out and so that I think those are good for, you know, being able to solve problems by synthesizing information in different areas. In this case information with different areas would be, you know, taking multiple equations that grab different concepts and putting them together~[I].” Overall, we see that Charlie’s educational goals for his class are to get students to be able to connect physics to their primary fields of study as well as real world scenarios that they encounter.

Molly’s educational goals are centered around her students developing the use of multiple tools. Molly teaches a physics class for biology students. For her class, she thinks it is important to leave with “a toolbox of skills[I].” This ‘toolbox’ is composed of “physics words and representations and like common usages of physics~[I]” that they can use when they “encounter a new situation[I].”  These tools should also “help them look at problems in a way that is not the same as what they get from biology and chemistry[I].” 

Our Fellows all have educational goals that align with the 3DL framework. More specifically, all of these goals are concerned with students learning the skills that scientist use. Using the principles of LGDD, these educational goals are used to help create the learning goals for their 3DL material. The next steps of LGDD have the Fellows aligning their course materials to their learning goals.

\subsubsection{Evaluation}

Our Fellows all take a different approach to their assessment design. Their different approaches are influenced by resources used throughout their departments and their current assessment practices. 

Ron talks a lot about a course management system that his department uses, and how it affects the design of his assessment items. More specifically, Ron wants to move away from the way that exams are typically done, but struggles to do so. One reason for this is because “we use [a course management system]. And it's really, really well-suited, well-designed for multiple choice problems~[I].” When it comes to formal assessment for Ron, he feels limited to using a multiple choice format. His formal exams are typically multiple-choice questions and “similar to the homework~[I].”  The course management system that Ron uses restricts him from using other types of  questions. Ron does not find multiple choice formatting conducive to the 3DL framework. “At least from going through the Fellowship so far, I didn't get the impression that it was… It was easy to have something that you could just have like four or five multiple choice questions that you could have as a 3D assessment~[I].”  

Charlie, our next Fellow, has already built his exams from his instructional materials and likes the concepts that they test. We see this when Charlie talks about how he has built his instructional material as well as his current exams. Charlie has already integrated clicker questions into his lecture that help his students on exams. He describes his clicker questions as opportunities for students to “see more of the types of logic that I want them to be able to do when they get in front of a test~[I].” Charlie also likes his current exam practices, and this changes the way that he views 3DL. One reason for this is because he “take[s] a fair amount of pride~[I]” in his assessment questions. More specifically, he “like[s] the concepts they target”, and “like[s] that, the way that I think they make the students think~[I].” Charlie’s feelings toward his current assessment practices make him unsure about changing them in order to integrate 3DL. 

Our final Fellow, Molly is currently in the process of aligning her instruction of scientific practices with their assessment. Molly’s exams have a group component where she is “testing [student’s] experimental understanding~[I].” this aligns closely with her want to assess scientific practices.

Ron, Charlie, and Molly develop their exams differently. Ron develops his exams using a course management system that his department uses. The use of this system limits Ron to the kind of questions that he can use on his exam. Charlie already uses a specific process of creating and modifying his exams to align with his current instruction, and Molly is in the process of integrating scientific practices into her assessments.

\subsubsection{Connection to Outlook}

Ron, Charlie, and Molly all reflect on their process of aligning their learning goals, instructional material, and their assessment. Ron and Charlie both articulate their reservations with their misalignment while Molly talks about the benefit of having her materials align with each other. 

Ron is unable to align his instruction with his assessment and this disconnect causes discomfort when it comes to continuing with 3DL.  Ron mentions that his exams and instruction don’t currently align, and this makes him doubt the effectiveness of bringing 3DL into his classroom. Ron brings up in the interview, “it's all fine and good if you have examples in class and if you give the homework like that, but if it's not, if it’s also not an exam question, then the students aren't going to really invest themselves in learning how to do those kinds of problems~[I].” Ron even notes needing help creating 3D material in the end of the year survey, “I would like to learn how to write 3DL assessments within a [course management system] framework. This will help me design new exam questions~[F] . Ultimately Ron viewed this as “one opportunity for the students to engage with material in this way and then they didn’t see it in this format again on the homework or the exam and I think it would really drive it home~[I].”

Charlie’s discomfort with 3DL comes from the way that he feels about his current material and his alignment between his instructional and assessment items. Charlie likes his current instructional and assessment practices. In particular he likes the concepts his test questions cover and thinks his instruction and assessment complement each other nicely. He states, “I have been orienting my class, I know what questions I like to ask~[I].” In the end, Charlie comes to the conclusion that “my default probably ought to be to show them questions and recitations that more closely connects to types of test questions~[I].” He also mentions that aligning his 3DL instruction to assessment is difficult due to “spen[ding] so much time and effort on [the recitation] that it never turned into an exam problem~[I].” He goes on to say “integrating the three dimensions into the test questions is significantly difficult and challenging, that it almost decouples the test from the teaching is that it sort of breaks that connection there~[I].” Ultimately, he sees creating the assessment item as an “afterthought~[I].”

Molly’s ability to align her exams with her instruction leave her feeling positively about her 3D material. Molly mentions how working with others inside of the Fellowship contributed to her success in aligning her assessment and instruction. In her interview she states, “I've like really enjoyed hearing from people and the different ways that they write exam questions and think about grading them~[I].” Molly also talks about these social interactions as “really helping me with my exam questions make them better using 3D, 3D dimensional learning~[I].” Molly specifically mentions working with Ruben (pseudonym), a Fellowship coordinator. She mentions that “[Ruben] would look at my exam questions, and he would help me see that, while implicitly there was a practice there, I wasn't explicitly asking students to show me their understanding of the practice[I].”The end results were very positive. Molly recalls, “my questions just got much cleaner and like students would produce what I really wanted them to do~[I].” 3DL assessments have allowed Molly to pay closer attention to what she was assessing.

Ron, Molly, and Charlie all mention alignment between instruction and assessment as a reason to either continue or discontinue their use of 3DL material. Ron and Charlie both note misalignment as a reason to stop the use of their material. On the other hand, Molly notes being able to assess her students on her instructional items as a reason to continue developing more material in the future.

\subsection{Classroom Structures and Instructional Practices}

Another prevailing theme is the different classroom structures and instructional practices that our Fellows take part in. These influences are mentioned in the transformation and instruction stages of PR\&A, and ultimately affect how our Fellows are able to reflect on their 3DL material. 

\subsubsection{Transformation}
All three of our Fellows teach courses with different structures. As they talked about designing material for their course, they all mention how the structure of their course influenced the ways in which they designed their material. 

Ron teaches a class with only a lecture component. The educational resources that Ron uses allows him to create a student led activity in the lecture format, Although Ron is able to create an activity for his lecture-based class, he says that the structure of his class made the design of the activity challenging. In an end of the year survey, Ron writes, “[Developing material that gets students to engage in 3DL] was challenging because of the “pull” of the traditional lecture style is surprisingly hard to resist~[F].” Although the “pull” of the lecture is hard to resist Ron creates a class activity for his 3DL unit.

Charlie teaches a class with lecture, lab, and recitation components. Therefore, Charlie chose where to place his 3DL activity. Due to the structure of the class, Charlie also had to decide where he would like to introduce 3DL into the classroom. A couple factors like time and class structure help Charlie decide where he should place the activity in his class.
Charlie sees “time is at a, at a real premium~[I]”, and because of this “spending, you know, 15 minutes, even 15 minutes doing the lecture activity would have been brutal~[I].” Presenting a 3D activity in the recitation seemed to make the most sense to him because of the nature of the class. Charlie goes on to say, “The recitation seems like the ideal place to put the sort of thing. That's the time where the students really have time to, you know, get involved and, and get their hands dirty~[I].” We see 3DL appearing within the recitation component of Charlie’s class due to the problem-solving environment and the time that the class provides.

Molly teaches a studio class, and the studio format of Molly’s class allows her to focus on scientific practices by exploring biological phenomena. In her interview, Molly does mention her experience teaching lecture-based class.  She mentions, “so I think it was a lot harder. And that in that setting to think about some of those practices as fitting like if I felt like I should belong in the lab~[I].” With the studio structure of her current class students are able to engage in practices during lab.

All three of our Fellows had to think about different environmental factors when designing their material. Ron had to think about a way that he could design a class activity in a course that is designed to support direct instruction, Charlie had to decide what component of his course that he wanted to design material for, and Molly felt like her current course structure compared to her previous course structure supports her material design. Overall, we notice that all three of our Fellows selected different course components to adapt their material based off of resources available to them.

\subsubsection{Instruction}
The placement of their material influences their instructional practices. We previously saw that each Fellow designed material for different components of their course. Due to their material placement we see that their participation in instruction is also diverse.

Ron prepared his material for his lecture-based course. This allows Ron to be present for the instruction of his material. Although Ron is always present for instruction, his instruction of his unit deviated from his typical instruction. During the instruction of Ron’s 3DL unit he was not the only instructor. Ron explains the instruction of the unit as the following: “when we did this in the lecture room, [my colleagues] and I were all [...] there, plus the TA for the class, plus the three LAs for the class~[I].” The role of the instructors for this unit focused mostly on classroom organization and group facilitation by “walking around the outside~[I]” to check on students. This deviates greatly from Ron's typical approach which is “just straight lecturing~[I].”

Due to Charlie creating material for his recitation component, Charlie’s instruction relied on himself as well as other instructors. Charlie was present for the instruction of his 3D material in the lecture and he “really like how this fit into the lecture~[I]”, but he was not able to be present for the instruction of the 3D recitation material. During the recitation different instructors are responsible for teaching the 3D material. In recitation, instructors give students a “problem to work out […] usually in teams of four and […] they've got a secondary instructor roaming the room to and hopefully an LA but not this semester roaming the room to […] help them through that~[M].” During the 3D unit, instructors decided to encourage “a bit of a divide and conquer approach~[M]” this was mainly done to “make things go more efficiently~[M].” Charlie was able to get this feedback from another Fellow who teaches his recitation although he typically doesn’t “feel like [he] need[s] to check in~[I].”

Molly produced material for her studio class, which has allowed her to develop her own instructional technique. Molly’s instruction of her material is through a process that she calls “applying and extending~[I].” In her interview Molly describes this process, “I present them a problem, a phenomenon of some sort. And, and I asked them to think about, like, what kinds of things would they like to know about it? And, and then they, they designed some experiment, or to figure out some of the stuff that they would like to know about it […] And then we spend the next day seeing how far those rules will take us~[I].”

All of our Fellows chose a different way to instruct their course materials. Their decisions for instruction were largely due to the structure of their course. Ron and Molly were both there for instruction, and Charlie was only present for some of the instruction of his materials. 

\subsubsection{Connection to Outlook}
Each Fellow developed materials based on their course structure, and therefore participated in instruction differently. Ultimately, we see that the different ways that they participated in instruction influences their overall thoughts about their developed course material.

Ron’s participation in instruction resulted in him developing a personal insight on the ways in which students responded to the material and how they felt while he taught the material. In particular we see that the way the classroom felt after implementing the 3D material in his classroom leaves Ron with an optimistic feeling about 3DL. Ron notes how there was a “palpable difference~[F]” in the room, and in his interview Ron states, “I know that [feeling] shouldn’t be the gauge […] but putting that aside, they certainly seem to enjoy it more and we enjoyed it as well~[I].” The impressions that Ron was able to form from instruction served as a main motivating factor to continue with the use of 3DL. When asked about his thoughts about continued use Ron states that he is “very enthusiastic about it~[I].” 

Charlie had to rely on student grades and the feedback his colleague provided. The feedback he got from his colleagues and the grades were neutral, but he formed a negative impression about continuing with 3DL. Charlie states that the problem seemed to be “a fairly big departure from the types of problems they’re used to seeing in recitation~[M].” But overall, the unit didn’t seem much different from other units in the class. Charlie expresses this as a “lateral step~[M]”, and describes student understanding as “the total amount they learned was no more no less~[I]” and their grades “being within the noise~[I].” 

When it comes to reflecting on his continued use of his material, Charlie does not think he is going to continue to use 3DL. In particular, Charlie is unsure what to do with the recitation problem. He states, “I'm not sure. Whether this is this should be tweaked, or whether it'd be more useful for them to do some more evolved conservation energy type calculations to get an idea of the utility of, of, of the method~[M].” This leads to uncertainty about continuing use of 3DL Charlie states, “I don't know. I mean, at this point, I think the easy answer is no~[I].”

Molly reflects on her method of applying and extending, mentioning that “students are constantly surprising me with things that I didn't think about. Either they're thinking about something I did think about it, but they're thinking about it in a new way that I haven't considered before [I].”

We see that our Fellows decisions about material placement affect their instruction and ultimately their feeling about continued use. Ron and Molly were both able to be present for the instruction of their material, and therefore reflect on how the class felt and how students were responding to the new material. On the other hand, Charlie was not available for the instruction of his material and spends his time reflecting on student grades rather than his perception of how the material went. These factors lead to Ron and Molly feeling positive about continued use, but leaves Charlie discontinuing the use of his material.

\subsection{Social Interactions}

We see that all three Fellows are influenced by social interactions. These interactions influence the transformation and evaluation stages of our Fellows design process. In particular we see that social interactions with other Fellows and students have a positive effect on our Fellows while interactions outside of the Fellowship tend to negatively affect our Fellows design process. 

\subsubsection{Transformation}
The interactions that all three Fellows have inside the Fellowship influence the ways in which they learn about new teaching ideas and supports them in their material development. All three of our Fellows mention interactions they have with colleagues inside of the Fellowship. Ron also comments on his experiences with colleagues outside of the Fellowship.

Ron experiences social interactions inside and outside of the fellowship that influence his material design. When talking about interactions with other Fellows Ron describes it as being “very helpful because it's certainly opened my eyes to what could be done in the large lecture room~[I].” Ron also mentions that he used “feedback [from other Fellows] about how to refine the question so they were closer to 3D.” This is contrasted by the way Ron talks about interactions with non-Fellows, Ron describes these faculty as tending “to have a more old school perspective on the teaching is just, there's some lecture slides, just give a good performance, the students are going to do how they do and this is really out of your control. And this is kind of the feedback that I get outside the Fellowship, to be honest~[I].” These types of interactions have stopped Ron from talking about his ideas to change features of his advanced lab course that he just began co-teaching with a colleague from outside of the Fellowship. When asked about talking to his new co-instructor about integrating his ideas of 3DL into the advanced lab classroom, Ron responds, “Uh, no, I probably should … I should give him, I should give the benefit of the doubt…~[I]” Overall, Ron feels supported by his colleagues in the Fellowship, but having a hard time finding support from other colleagues outside of the Fellowship.

For Charlie, a lot of the design decisions that went into creating his 3DL unit came from working with another Fellow. While creating his 3DL unit Charlie picked the content of the unit so it could be a collective effort with another Fellow. Charlie describes this process as, “so we figured that these are very similar concepts. And so we could put together a teachable unit that can be used with minor modifications in both, both contexts~[F].”

Molly also talks about the benefit of working with Fellows from other disciplines helped Molly with her material design. During the Fellowship Molly and other Fellows designed a cross disciplinary problem for students. Molly recalls this experience as “a cool thing that wouldn't have come about if I had not been in a group of people with Biology and Chemistry and like all of us together[I].” 

Overall we see that interactions with others in the Fellowship provide our Fellows with support for their material design through the use of sharing ideas and providing feedback. Ron also mentions his experiences interacting with colleagues outside of the Fellowship. These interactions do not provide Ron with the support to implement new ideas as he is often told to stick to the ways instruction has been done in the past. 

\subsubsection{Evaluation}
When it comes to developing and thinking about assessment, two of our Fellows mention they were influenced by social interactions either inside or outside of the Fellowship. 

When it comes to 3DL assessment, Ron thinks about his students, his colleagues inside of the Fellowship, and his colleagues outside of the Fellowship. When it comes to his students, Ron is “very comfortable [giving a 3D exam]~[I]” and “in terms of my students I think it would be great for them to do it and I think they would rise to the occasion~[I].” Ron also mentions how his assessment practices are influenced by his colleagues. In particular, Ron is “very comfortable [giving a 3D exam]~[I]” if he was talking to other Fellows, but Ron knows that his colleagues outside of the Fellowship “would recommend against it.” Overall, the social interactions with people outside of the Fellowship hinder the spread of 3DL throughout Ron’s department. 

Molly found support for assessment design and practice within the Fellowship. This support helped her generate new ideas for exams as well as generating questions that align more closely with 3DL. In her interview, she states, “I've like really enjoyed hearing from people and the different ways that they write exam questions and think about grading them~[I].” Molly also talks about these social interactions as “really helping me with my exam questions make them better using 3D, 3D dimensional learning~[I].” Molly specifically mentions working with Ruben (pseudonym), a Fellowship coordinator. She mentions that “[Ruben] would look at my exam questions, and he would help me see that, while implicitly there was a practice there, I wasn't explicitly asking students to show me their understanding of the practice[I].” The end results were very positive. Molly recalls, “my questions just got much cleaner and like students would produce what I really wanted them to do~[I].” 3DL assessments have allowed Molly to pay closer attention to what she was assessing.

Both Ron and Molly mention the benefit of adapting 3DL assessment items into their classroom. Ron remains apprehensive about creating 3DL exams due to interactions outside of the Fellowship, but unlike Ron, Molly was able to find a lot of support when redesigning her exams to assess scientific practices.

\subsubsection{Connection to Outlook} 
As outlined in other sections of our Results we see that both Ron and Molly are optimistic when it comes to continued use of 3DL in their course and instruction. Ron’s decision to continue use is largely due to the interactions he has had with his students. As mentioned earlier Ron talks about the “palpable difference” he felt while teaching as a motivating factor to move forward with 3DL. Although we do see interactions that Ron has outside of the Fellowship as hindering the spread of 3DL. In particular, his hesitation to share ideas with his co-instructor in his Advanced lab class. Molly’s plans for continued use not only come from interacting with her students, but also being able to accurately assess and instruct scientific practices, a skill she was able to develop by working with other Fellowship colleagues. In both cases we see the social interactions that the Fellows inside of the Fellowship as contributing to their continued use of 3DL.

\section{Discussion\label{discussion}}



    
    
    
    
    
    

In our study we investigated influences on faculty uptake utilizing PR\&A as our analytic framework. We explored the PR\&A of three cases in order to determine the influences on the plans for continued use of 3DL. Our cross case analysis of Ron, Charlie, and Molly has pointed to assessment alignment, participation in instruction, and social interactions in and outside of the Fellowship influence the ways faculty can take up ideas from a FLC. 

Our first research question was concerned about different factors that push Fellows towards or away from the continued use of 3DL. Alignment of ideas in the comprehension, evaluation and instructional stages of material design promotes continued use of 3DL. More specifically, the alignment of assessment and instructional practices lead to a positive outlook on 3DL. We also found that participation in the instruction of classroom materials have a positive effect on the Fellow's outlook on 3DL. 

Secondly, we explored the ways in which the FLC supports Fellow’s adoption and plans for continued use. We discover that the FLC provided Fellows with a space to share new ideas, learn from other’s ideas, and work with other Fellows on material design. The FLC also spent time talking about the importance of aligning instruction and assessment; although this does not help Ron and Charlie to align their instruction and assessment, it does influence how they view the utility of their 3DL unit.

\subsection{Factors Associated with Continued or Discontinued Use} 

Our study revealed several factors that are associated with Fellows deciding to continue or discontinue use of their 3DL materials. These factors are associated with the ability to align assessment and instruction and the participation in the instructional component of the material.  

Alignment of ideas in the comprehension, evaluation and instructional stages of material design promotes continued use of 3DL. More specifically, the alignment of assessment and instructional practices lead to a positive outlook on 3DL. 

The Fellowship is designed around the premise that changing assessment leads to change in instruction~\cite{laverty2014}, and our research suggests that there is a correlation between the ability to align instruction to assessment and continued use of 3DL. More specifically we found that assessment design is influenced by the current state of assessments tools/technology, and how well it can be integrated into instruction. One challenge that we found is Ron’s trouble with the course management system that his department uses. This reiterates the work of Zohrabi that finds faculty’s motivations for adopting RBIs can be negatively impacted due to the use of certain textbooks or online platforms~\cite{zohrabi2020}. But, unlike Zohrabi we see that Ron’s use of the course management system affects his use of assessment rather than motivation to change it. There is a lot of work out there that outlines ways for faculty to align their instruction with assessment~\cite{krajcik2008, mislevy2011, wiggins1998}, but there is little work that investigates the challenges faculty encounter when aligning their materials or developing assessment items. 

Our study reveals participation in instruction and the ability to align assessment to instruction all positively impact our Fellow’s continued use of 3DL. More specifically we find faculty participation and classroom structure influence how faculty feel about the change that they are making. Therefore, it is important for faculty to collect evidence of student learning as well as participating in teaching experiences in order to use their intuition. 

The structure of the classroom, and faculty participation in instruction influence Fellow’s plans for the continued use of 3DL. Structures of classrooms are often out of the control of instructors, and this can impact how faculty interact with their classrooms. More specifically, classrooms that rely on more than one instructor have an added barrier as other instructors might not be aware of the main instructor’s goals~\cite{west2013}. Other research has found that faculty often use their intuition and general feelings to see if a technique worked~\cite{zohrabi2020}, and that the types of interactions instructors have with students during instruction also affect their use of pedagogical practices~\cite{tinnell2019}. This all goes to support Turpen, Dancy, and Henderson’s claim that faculty are more convinced by their own experiences rather than with data~\cite{turpen2016}. 

Lastly, the social interactions that our Fellows have in and outside of the Fellowship effect the use of the materials. Social interactions include interactions with students, Fellows, and colleagues outside of the Fellowship. Interactions with students give instructors a feel for their material and reiterate the need for Fellows to participate in instruction. Interactions within the Fellowship have allowed Fellows to receive valuable feedback and to share ideas about bringing 3DL into the classroom and aligning assessment and instructional materials. However, interactions with colleagues outside of the Fellowship are not centered around sharing ideas about 3DL.

\subsection{The Role of FLC in Promoting Continued Use} 

Faculty Learning Communities are created in order to build a community of faculty that are open to discussing and sharing ideas about teaching practices. Throughout all three cases we see that FLC succeeded in creating a space in which faculty are open to receiving and discussing new ideas and feedback. Our work suggests that FLCs also need to help Fellows align their instruction and assessment, encourage Fellows to experience teaching their new materials, and spread their ideas to their colleagues and throughout their department.

Alignment between instructional material and assessment items are correlated with the exclusion of 3DL material. In order to prevent discontinued use of their design ideas due to assessment practices, FLC facilitators should include discussions on how to design assessment items and address common issues such as formatting.  Although the Fellowship spent time talking about assessment design and the process of aligning assessment with instruction, it did not dedicate time to understand the logistics of assessment design and the current state of their assessments beyond talking about what it is that they want to assess. Professional development has recently begun to make the effort to integrate the topic of assessment into programs~\cite{kinzie2019, caudle2018, hong2018}, but there needs to be more research on the ways in which faculty develop and use assessment items. FLCs should address the topics surrounding current assessment development as a way to help faculty adopt new instructional practices. 

Participation in instruction is a valuable experience, and in our cases, we see this directly affect buy in to 3DL. FLC designers need to understand that data does not always drive change in order to encourage instructors to participate in instruction and use their intuition as valuable data. This sentiment is echoed by Henderson, Dancy, and Niewiadomska’s findings that support provided from professional development leaders rarely includes observation and feedback during implementation~\cite{henderson2012}. One solution for this could be the use of peer-to-peer observations~\cite{tinnell2019, onodipe2020} or by providing faculty with opportunities to participate in and reflect on their own experience with new material as well as the ability to learn from student outcomes

Lastly, our Fellows come from different departments, and this has an effect on the spread of ideas throughout the department. In the future, Faculty Learning Communities need to focus on the departmental culture that they are a part of and be attentive to the way these cultures could affect the spread of new ideas. Gehrke and Kezar find that faculty members can enact departmental and institutional change~\cite{gehrke2017}. FLCs that acheive change throughout a department incorporate enacting change are large peer groups and developing leadership skills from participation in the community~\cite{gehrke2017}. This research supports our research findings and the need for FLCs need to work towards supplying participants with the tools and resources that will help them grow their community of peers as well as promote change within their department.

Our work suggests that FLCs should spend time talking about assessment, encouraging participation in instruction, and supply the necessary tools to help promote change. We also recommend that the research community investigates the ways in which faculty use and design assessment items.

\section{Conclusion \label{conclusion}}

Our research has found that that the alignment of assessment and instruction as well as participation in instruction contribute to the continued use of 3DL. Alignment of assessment and instruction can be influenced by course structures, course management systems, and social interactions with other Fellows. Participation in instruction can also be influenced by the course structure of the class, and allows instructors to be able to use their experience and feelings about their material as reasons for continued use.

Along with our results from our research we found that the process of assessment design from the perspective of the practitioner is also underrepresented in the literature. As a research community we need to further investigate how faculty use and design assessments.

\section{Acknowledgements}
A special thanks to the Kansas State University Physics Department, the faculty involved in interviews, and my 3DL4US colleagues especially Paul Bergeron and Marcos D. Caballero. This material is based upon work supported by the National Science Foundation under Grant No. 1726360.

\bibliographystyle{apsrev}

\bibliography{LydsBib}

\end{document}